\documentclass[journal,10pt]{IEEEtran}

\usepackage{textcomp}
\usepackage{comment}
\usepackage{cite}
\usepackage{graphicx}
\usepackage{epsfig,graphics,subfigure,psfrag,amsmath,amssymb}
\usepackage{amsmath}
\usepackage{slashbox}
\usepackage{multirow}
\usepackage{amssymb}
\usepackage{algorithm}
\usepackage{algorithmic}
\usepackage{bm}
\usepackage{hyperref}
\usepackage{url}
\usepackage{breakurl}
\usepackage{graphicx}
\usepackage{mathrsfs}
\begin{document}
%
\title{Enhanced Random Access and Beam Training for mmWave Wireless Local Networks with High User Density}
%
%
%

\author{Pei~Zhou,
        Xuming~Fang,~\IEEEmembership{Senior Member,~IEEE,}
        Yuguang~Fang,~\IEEEmembership{Fellow,~IEEE,}
        Yan~Long,
        Rong~He,
        and~Xiao~Han

\thanks{P. Zhou, X. Fang, Y. Long and R. He are with Key Lab of Information Coding \& Transmission, Southwest Jiaotong University, Chengdu 610031, China (e-mail: peizhou@my.swjtu.edu.cn; xmfang@swjtu.edu.cn; yanlong@swjtu.edu.cn; rhe@swjtu.edu.cn). }

\thanks{Y. Fang is with the Department of Electrical and Computer Engineering, University of Florida, PO Box 116130, Gainesville, FL 32611, USA. (e-mail: fang@ece.ufl.edu).}

\thanks{X. Han is with the CT lab, Huawei, Shenzhen 518129, China (e-mail: tony.hanxiao@huawei.com).}}

{}

\maketitle

\begin{abstract}
As low frequency band becomes more and more crowded, millimeter-wave (mmWave) has attracted significant attention recently. IEEE has released the 802.11ad standard to satisfy the demand of ultra-high-speed communication. It adopts beamforming technology that can generate directional beams to compensate for high path loss. In the Association Beamforming Training (A-BFT) phase of beamforming (BF) training, a station (STA) randomly selects an A-BFT slot to contend for training opportunity. Due to the limited number of A-BFT slots, A-BFT phase suffers high probability of collisions in dense user scenarios, resulting in inefficient training performance. Based on the evaluation of the IEEE 802.11ad standard and 802.11ay draft in dense user scenarios of mmWave wireless networks, we propose an enhanced A-BFT beam training and random access mechanism, including the Separated A-BFT (SA-BFT) and Secondary Backoff A-BFT (SBA-BFT). The SA-BFT can provide more A-BFT slots and divide A-BFT slots into two regions by defining a new `E-A-BFT Length' field compared to the legacy 802.11ad A-BFT, thereby maintaining compatibility when 802.11ay devices are mixed with 802.11ad devices. It can also reduce the collision probability in dense user scenarios greatly. The SBA-BFT performs secondary backoff with very small overhead of transmission opportunities within one A-BFT slot, which not only further reduces collision probability, but also improves the A-BFT slots utilization. Furthermore, we propose a three-dimensional Markov model to analyze the performance of the SBA-BFT. The analytical and simulation results show that both the SA-BFT and the SBA-BFT can significantly improve BF training efficiency, which are beneficial to the optimization design of dense user wireless networks based on the IEEE 802.11ay standard and mmWave technology.
\end{abstract}


\IEEEpeerreviewmaketitle

\section{Introduction}

\IEEEPARstart{W}{ith} the popularity of ultra-high definition video (UHD Video), virtual reality (VR) equipments and future fifth generation (5G) mobile communication systems, the low frequency band becomes more and more congested. Hence, millimeter-wave (mmWave) band has attracted much attention due to large spectral resource availability. The Federal Communications Commission (FCC) recently released 3.85 GHz licensed spectrum (i.e., 27.5-28.35 GHz, 37-38.6 GHz and 38.6-40GHz) and 7 GHz unlicensed spectrum (i.e., 64-71 GHz) for wireless cellular systems \cite{ref1}. In addition, FCC continues to seek for opinions on above 95 GHz frequency band to address the spectral resource requirements. This is an important opportunity for the development of ultra-high speed wireless communications. Meanwhile, Verizon recently released V5G standard operating at 28-40 GHz frequency band \cite{ref2}, and IEEE released the Wireless Gigabit Alliance standard(WiGig \cite{ref3}, which is unified with the Wi-Fi Alliance \cite{ref4}), IEEE 802.11ad (TGad) \cite{ref5}, \cite{ref6}, IEEE 802.15.3c (TG3c) \cite{ref7} and WirelessHD (WiHD) \cite{ref8}, operating at 60GHz unlicensed band to meet the demand of future high speed wireless communications. Consequently, 3GPP 5G standard workgroup has begun to discuss how to adopt mmWave band in cellular systems. Although mmWave band has rich unlicensed spectrum available, the propagation suffers from serious path loss \cite{ref9}, \cite{ref10}. In order to compensate for serious path loss and support relatively long-range transmissions, relays and directional beamforming technologies are the key enablers. Aiming at extending the communication range and guarantee the end-to-end performance, Yang et al. in \cite{ref11} proposed a multi-hop 60 GHz wireless network for outdoor communication where multiple full-duplex buffered relays are used. Relays can also be used to establish the communication links where line-of-sight path is unavailable. To analyze the performance of average throughput and outage probability, Yang et al. in \cite{ref12} proposed a maximum throughput path selection algorithm to select the optimal path that maximizes the throughput. Since narrow beams may cause frequent beam switching, we proposed a robust and high throughput beam tracking scheme in mobile mmWave communication systems to balance both throughput and beam handoff probability \cite{ref13}. In dense user scenarios, the beams that serve different users may transmit in the same path, and therefore imperfect orthogonal beams may cause severe interference. In \cite{ref14}, Xue investigated the interference of nonorthogonal beam, then developed dynamic beam switching and static beam selection schemes to coordinate the transmitting beams effectively.

Beam alignment is a key design issue to achieve high speed and high quality data transmissions when taking advantage of beamforming antenna gain. Therefore, communicating nodes have to trigger beamforming (BF) training processes before directional data communications \cite{ref14}. To address this issue, various BF training methods have been proposed. An exhaustive beam searching method to find the best pair of transmit and receive beams was proposed in \cite{ref15}, which can achieve beam alignment and provide high beamforming gain. However, the training processes in \cite{ref15} are very inefficient. An alternative method proposed in \cite{ref16} and adapted by IEEE 802.15.3c standard \cite{ref17} employs a binary search BF training algorithm based on layered multi-resolution beamforming codebook to reduce training time. However, this method focuses only on efficient beamforming training for point-to-point communications without providing any solution for multi-user communication scenarios. Noh et al. in \cite{ref18} considered the design of multi-resolution beamforming sequences to enable an mmWave communication system to quickly find the dominant channel direction for a single path channel. In \cite{ref19}, a new BF training technique called beam coding was proposed, which not only shows the robustness in non-line-of-sight environments, but also provides very flat power variations within a packet. In contrast to the IEEE 802.11ad standard, the proposed scheme may lead to large dynamic range of signals due to the variations of beam angles within a training packet. In \cite{ref20}, an efficient and low-complexity codebook-based BF training technique was proposed for short-range indoor communications, which is based on the implementation of the Nelder-Mead simplex algorithm iteratively and recursively. The proposed scheme can achieve the similar beam selection function compared to the exhaustive BF training algorithm in \cite{ref15}. Song et al. in \cite{ref21} discussed an mmWave system employing dual-polarized antennas which will reduce the time for beam training. From another point of view, most of the previous studies focused on the analog beamforming with the objective of improving the average signal to noise ratio (SNR). In contrast, Li et al. in \cite{ref22} proposed to minimize the mean square error (MSE) of the baseband equalized signal. The IEEE 802.11ad standard defines a BF training scheme that consists of Sector Level Sweep (SLS) phase and Beam Refinement Protocol (BRP) phase \cite{ref10}, \cite{ref23}. A typical SLS phase consists of several sub-phases, and the most important sub-phase of SLS is the association beamforming training (A-BFT). It is used by stations (STAs) to access channel and train their transmit antenna sectors for communications with personal basic service set (PBSS) control point/access point (PCP/AP). During the A-BFT, a large number of STAs randomly and independently select a slot among the limited number of A-BFT slots (i.e., at most 8 slots) to access channel and perform BF training which will suffer a high probability of collisions and thereby cause poor performance, especially in dense user scenarios. In such scenarios, Kim et al. proposed a load balancing approach in mmWave wireless local area networks (WLANs) to mitigate collisions by spreading out the random-access attempts over time \cite{ref24}. However, it will need to extend the BF training time. An efficient A-BFT beam training procedure that allows different STAs to transmit training frames simultaneously over multiple channels for dense user scenarios was proposed in \cite{ref25}. However, it will make the training phases after A-BFT very complicated. Therefore, it is crucial to design time efficient BF training methods to improve the BF training performance.

This paper dedicates to address how to deal with high collision probability and low BF training efficiency of the IEEE 802.11ad in dense user scenarios. We first propose a Separated A-BFT (SA-BFT) mechanism that can provide more A-BFT slots, which will significantly alleviate the collisions in A-BFT phase. Then we present a Secondary Backoff A-BFT (SBA-BFT) mechanism to further reduce the collision probability in the A-BFT phase. In order to guarantee the compatibility with the IEEE 802.11ad standard, the SA-BFT mechanism separates the Directional Multi-Gigabit STAs (DMG STAs) in the IEEE 802.11ad and the Enhanced DMG STAs (EDMG STAs) in the next generation of mmWave WLAN standards (i.e., the IEEE 802.11ay) into two different A-BFT slot regions while EDMG STAs can perform the SBA-BFT in the second A-BFT slot region.

The main contributions of this paper are summarized as follows:

1)	We propose an SA-BFT mechanism to provide more A-BFT slots compared with the IEEE 802.11ad standard, so that the collision probability during BF training phase can be reduced, and the compatibility with the IEEE 802.11ad standard can be maintained.

2)	Based on the SA-BFT mechanism, we develop an SBA-BFT mechanism to further reduce the collision probability in the A-BFT phase. By designing the secondary backoff window, the priority of failed STAs can be promoted and the timeliness of BF training can be improved.

3)  We present a three-dimensional Markov chain model to analyze and verify the performance of the SBA-BFT mechanism.

The remainder of this paper is organized as follows. In Section II, we describe the system model and BF training process of the IEEE 802.11ad. In Section III, the SA-BFT mechanism is proposed and the simulation results of performance comparison between the SA-BFT and the legacy IEEE 802.11ad A-BFT are given. Based on the SA-BFT, we further propose the SBA-BFT mechanism and set up a three-dimensional Markov chain model for the SBA-BFT in Section IV. Simulation studies of the SA-BFT and SBA-BFT are carried out in Section V and finally Section VI concludes the paper. In order to improve the readability of the paper, Table I summarizes the main acronyms used throughout the paper.

\begin{table}[!htbp]
\normalsize
\caption{SUMMARY OF MAIN ACRONYMS.}
\centering
\begin{tabular}{|c|c|}
\hline
\textbf{Acronyms} & \textbf{Definition} \\
\hline
ATI & Announcement Transmission Interval \\
\hline
A-BFT & Association Beamforming Training \\
\hline
BC & Beam Combining \\
\hline
BF & Beamforming \\
\hline
BI & Beacon Interval \\
\hline
BRP & Beam Refinement Protocol \\
\hline
BTI & Beacon Transmission Interval \\
\hline
CBAP & Contention-Based Access Period \\
\hline
CCA & Clear Channel Assessment \\
\hline
DCF & Distributed Coordination Function \\
\hline
DMG & Directional Multi-Gigabit \\
\hline
DTI & Data Transfer Interval \\
\hline
EDMG & Enhanced DMG \\
\hline
I-TXSS & Initiator Transmit Sector Sweep \\
\hline
MID & Multiple Sector Identifier \\
\hline
OI & Overload Indicator \\
\hline
PBSS & Personal Basic Service Set \\
\hline
PCF & Point Coordination Function \\
\hline
PCP/AP & PBSS Control Point/Access Point \\
\hline
R-TXSS & Responder Transmit Sector Sweep \\
\hline
SA-BFT & Separated A-BFT \\
\hline
SBA-BFT & Secondary Backoff A-BFT \\
\hline
SBIFS & Short Beamforming Inter Frame Spacing\\
\hline
SLS & Sector Level Sweep \\
\hline
SNR & Signal Noise Ratio \\
\hline
SP & Scheduled Service Period \\
\hline
SSW frame & Sector Sweep frame \\
\hline
STA & Station \\
\hline
\end{tabular}
\end{table}

\section{System Model}

\subsection{Network Topology}
In the IEEE 802.11ad standard, PBSS is composed of one PCP/AP and $N$ ($1 \le N \le 254$) non-PCP/non-AP DMG STAs as shown in Fig. 1. PCP/AP is the centralized control point of the PBSS, which is responsible for BF training, scheduling and channel access of the entire PBSS \cite{ref23}.

\begin{figure}[!htbp]
  \begin{center}
    \scalebox{0.48}[0.48]{\includegraphics{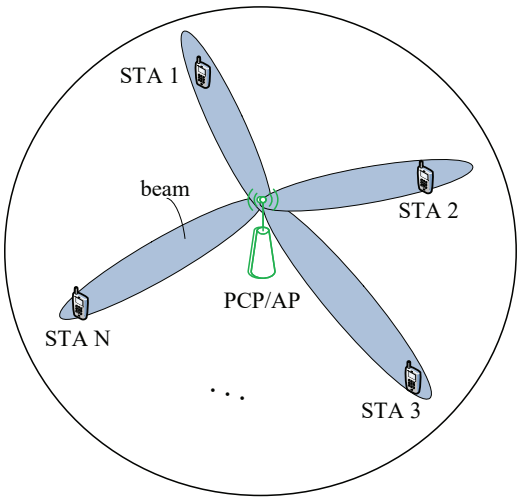}}
    \caption{IEEE 802.11ad PBSS topology.}
  \end{center}
\end{figure}

\subsection{IEEE 802.11ad Beacon Interval}
In the IEEE 802.11ad standard, time domain is divided into many Beacon Intervals (BIs), as shown in Fig. 2. One BI consists of four parts: Beacon Transmission Interval (BTI), A-BFT, Announcement Transmission Interval (ATI) and Data Transfer Interval (DTI) \cite{ref23}. In BTI, PCP/AP performs Initiator Transmit Sector Sweep (I-TXSS). In A-BFT, STAs mainly perform Responder Transmit Sector Sweep (R-TXSS). In ATI, PCP/AP allocates transmission opportunities of DTI for STAs. DTI is used for data transmission, which usually includes several Contention-Based Access Periods (CBAPs) adopting Enhanced Distributed Channel Access (EDCA), and Scheduled service Periods (SP) which is scheduled by the Quality-of-Service (QoS) AP or the PCP adopting Point Coordination Function (PCF) \cite{ref26}.

\begin{figure}[!htbp]
  \begin{center}
    \scalebox{0.62}[0.62]{\includegraphics{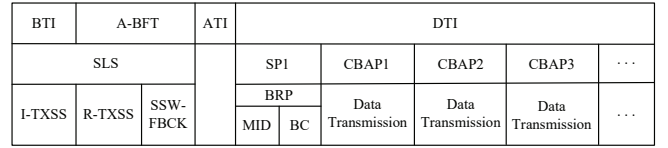}}
    \caption{Components of IEEE 802.11ad Beacon Interval.}
  \end{center}
\end{figure}

\subsection{IEEE 802.11ad Beamforming Training}
The IEEE 802.11ad BF training process consists of SLS and BRP, as shown in Fig. 2. The transmit beams of PCP/AP and non-PCP/non-AP DMG STAs are trained in SLS phase. BRP comprises two sub-phases: Multiple sector Identifier (MID) and Beam Combining (BC). In this paper, we focus on the training process of SLS. The detailed training process of BRP can be found in \cite{ref27}, \cite{ref28}, which is beyond the scope of this paper.

\begin{figure}[!htbp]
  \begin{center}
    \scalebox{0.8}[0.8]{\includegraphics{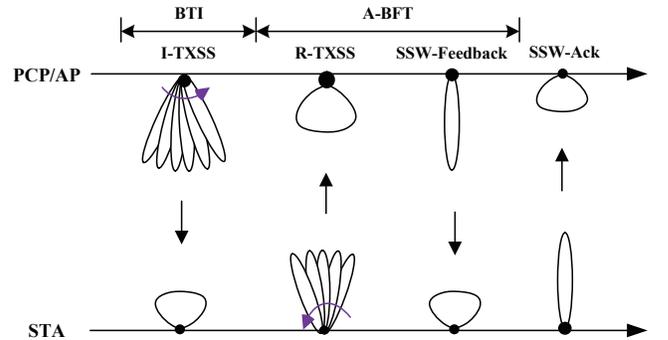}}
    \caption{BF training process of SLS.}
  \end{center}
\end{figure}

As shown in Fig. 3, in SLS, PCP/AP transmits DMG Beacon frames with different transmit beams to perform I-TXSS. Each STA receives DMG Beacon frames in quasi-omni mode. In A-BFT phase, each DMG STA performs R-TXSS by transmitting Sector Sweep (SSW) frames in different sectors, and the SSW frames contain the best transmit beam ID of PCP/AP. PCP/AP receives SSW frames in quasi-omni mode. Then during the SSW-Feedback sub-phase of A-BFT, PCP/AP employs its best transmit beam to feedback the best transmit beam ID of every successful STAs, respectively. It is worth noting that SSW-Ack is not mandatory. The A-BFT structure is shown in Fig. 4, in which, at most 8 A-BFT slots appear in IEEE 802.11ad A-BFT phase (indicated by the 3 bits `A-BFT Length' field) \cite{ref23}. It can be seen from Fig. 4 that slot 7 is selected by STA B and STA C simultaneously, thus both of them may fail on BF training, and they have to redo their training in the next BI.

\begin{figure}[!htbp]
  \begin{center}
    \scalebox{0.61}[0.61]{\includegraphics{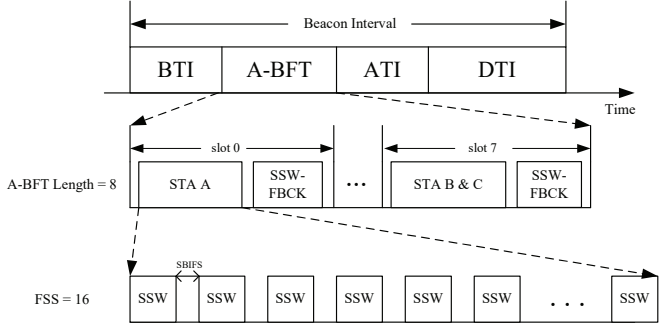}}
    \caption{A-BFT structure (take `A-BFT Length' = 8, FSS = 16 as an example).}
  \end{center}
\end{figure}

According to the reasonable settings of the IEEE 802.11ad standard in dense user scenarios, in which `A-BFT Length' = 8 \cite{ref23}, we use MATLAB to evaluate the performance of the IEEE 802.11ad A-BFT. The simulation results are shown in Fig. 5, where the horizontal axis stands for the number of STAs participated in A-BFT training contention, and the vertical axis stands for the average number of STAs successfully performed BF training. As it can be observed from Fig. 5, when the number of STAs participated in the A-BFT training is about eight, the number of successful STAs will reach the maximum value three. Then, the number of successful STAs decreases as the number of STAs increases due to the increased collision probability. According to the above analysis, we focus on the problem of high collision probability and low training efficiency of A-BFT phase in dense user scenarios.

\begin{figure}[!htbp]
  \begin{center}
    \scalebox{0.49}[0.49]{\includegraphics{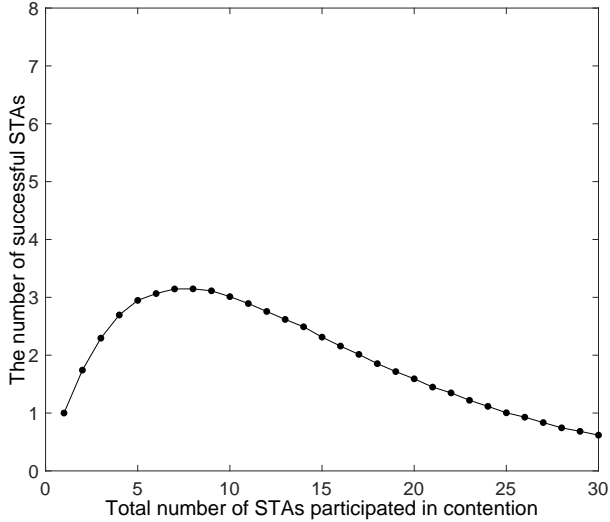}}
    \caption{Number of successful STAs in IEEE 802.11ad A-BFT.}
  \end{center}
\end{figure}

\section{Separated A-BFT Mechanism}
In future mmWave wireless networks, the DMG STAs in the IEEE 802.11ad standard and the EDMG STAs in the IEEE 802.11ay standard will co-exist for a relatively long time \cite{ref29}, \cite{ref30}, since the working mechanism of DMG STAs will keep unchanged. Therefore, for the future dense user scenarios, we propose an SA-BFT mechanism that can provide more A-BFT slots for EDMG STAs than that of the IEEE 802.11ad.

\subsection{Frame Structure Design of SA-BFT}
The structure of the Beacon Interval Control element of the DMG Beacon frame is shown in Fig. 6 \cite{ref23}, where `A-BFT Length' field indicates the A-BFT slot region. DMG STAs can randomly select a value between 0 to `A-BFT Length-1'. We occupy part of the reserved 4 bits (i.e., B44-B47) of Beacon Interval Control element (i.e., B45-B47) as an indicator (`E-A-BFT Length' field) of the number of A-BFT slots extended for EDMG STAs.

\begin{figure}[!htbp]
  \begin{center}
    \scalebox{0.51}[0.51]{\includegraphics{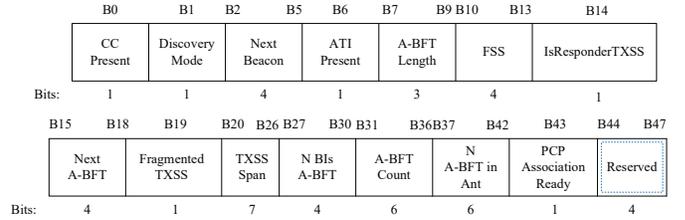}}
    \caption{Beacon Interval Control element of the DMG Beacon frame.}
  \end{center}
\end{figure}

\subsection{Signaling Design of SA-BFT}
In the SA-BFT, the start point of EDMG STAs in the A-BFT phase is A-BFT slot 0, and the A-BFT slots length of EDMG STAs is set to `A-BFT Length + E-A-BFT Length'. In the BTI phase, after an EDMG STA receives and demodulates the DMG Beacon frame, the A-BFT slot region they can randomly select from is uniformly distributed in [0, A-BFT Length + E-A-BFT Length-1]. As shown in Fig. 7, A-BFT slots for EDMG STAs are extended and redefined.

\begin{figure}[!htbp]
  \begin{center}
    \scalebox{0.62}[0.62]{\includegraphics{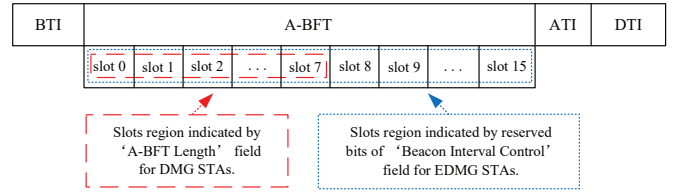}}
    \caption{Extended and redefined A-BFT for EDMG STAs.}
  \end{center}
\end{figure}

By setting the `E-A-BFT Length' to 8 (i.e., B45 B46 B47 = 111b), we can extend the original 8 A-BFT slots to 16 A-BFT slots. For the traditional A-BFT scheme, all EDMG STAs have to compete for the limited number of A-BFT slots indicated by `A-BFT Length' field. On the other hand, if the SA-BFT is executed, EDMG STAs shall have additional `E-A-BFT Length' A-BFT slots to compete for. To the best of our knowledge, there are no proper analytical models to determine the number of successful STAs. In that case, we design a corresponding algorithm (i.e., Algorithm 1) to obtain the number of successful STAs. A similar approach is given in \cite{ref31}. As shown in Fig. 8, when the number of STAs (to simplify the analysis, assume all STAs are EDMG STAs) increases from 1 to 30, the number of successful STAs in the SA-BFT outperforms that of the IEEE 802.11ad A-BFT. Obviously, the SA-BFT can increase the successful probability of A-BFT and improve BF training efficiency significantly in dense user scenarios. With additional slots provided, there are more choices for STAs to be randomly selected from. Thus, the collisions can be reduced and the number of successful STAs in A-BFT will be increased. As can be seen from Fig. 8, the more the additional slots are, the better the performance is.

\begin{figure}[!htbp]
  \begin{center}
    \scalebox{0.49}[0.49]{\includegraphics{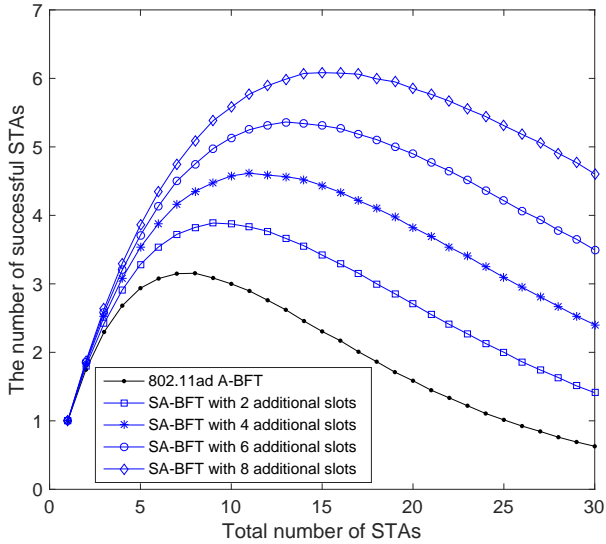}}
    \caption{The performance of SA-BFT with different number of additional slots.}
  \end{center}
\end{figure}

\subsection{Compatibility Design}
As shown in Fig. 9, the start point of ATI in the IEEE 802.11ad standard can be set by the Start Time field of the Next DMG ATI element of a DMG Beacon frame \cite{ref23}. In the SA-BFT mechanism, DMG STAs can avoid performing ATI immediately without waiting for EDMG STAs by adjusting the Start Time field of the Next DMG ATI element to a longer time, which equals to `A-BFT Length + E-A-BFT Length'. Thus, it is possible to guarantee that DMG STAs and EDMG STAs start ATI simultaneously. The SA-BFT procedure could be summarized in Algorithm 1.

\begin{figure}[!htbp]
  \begin{center}
    \scalebox{0.9}[0.9]{\includegraphics{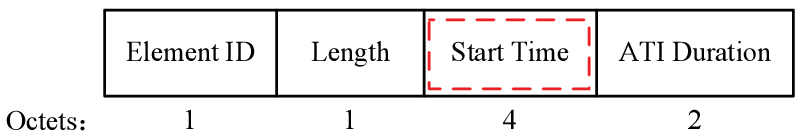}}
    \caption{Next DMG ATI element of DMG Beacon frame.}
  \end{center}
\end{figure}

\begin{algorithm}[!htbp]
  \caption{SA-BFT Algorithm.}
  \begin{algorithmic}[1]
    \STATE PCP/AP estimates the number of DMG STAs and EDMG STAs, and determines the value of `A-BFT Length' and `E-A-BFT Length' based on historical training information.

     \vspace{0.5em}
     \fbox{\shortstack[l]{Adjust the Start Time field of the Next DMG ATI ele- \\ ment to a longer time which equals to `A-BFT Length \\ + E-A-BFT Length'.}}
     \vspace{0.3em}

    \STATE PCP/AP transmits DMG Beacon frames to perform I-TXSS training.
    \STATE DMG STAs determine the A-BFT slot region according to `A-BFT Length' field of the received DMG Beacon frames.
    \STATE EDMG STAs determine the A-BFT slot region according to `A-BFT Length' field and `E-A-BFT Length' field of the received DMG Beacon frames, where the start point is A-BFT slot 0 and the length is `A-BFT Length + E-A-BFT Length'.
    \STATE Each of the DMG STAs and EDMG STAs randomly selects one A-BFT slot from the corresponding regions respectively to perform A-BFT beam training.
  \end{algorithmic}
\end{algorithm}

The proposed SA-BFT can extend the 802.11ad A-BFT slots to more A-BFT slots, which can alleviate the collision problem of A-BFT phase in dense user scenarios. Moreover, it can make DMG STAs and EDMG STAs compete for different A-BFT slot regions to maintain compatibility. By setting the start point of EDMG STAs to `A-BFT Length', we can adopt the SBA-BFT (proposed in section IV) mechanism to further reduce the collision in ultra-dense user scenarios. Since DMG STAs and EDMG STAs will compete for two non-overlapping A-BFT slot regions as shown in Fig. 10, there is no interference between them.

\begin{figure}[!htbp]
  \begin{center}
    \scalebox{0.62}[0.62]{\includegraphics{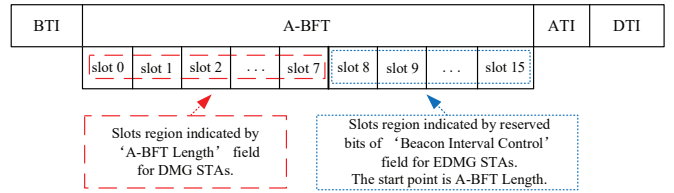}}
    \caption{The SA-BFT divides A-BFT slots into two non-overlapping regions.}
  \end{center}
\end{figure}

\section{The Secondary Backoff A-BFT Mechanism}
Although the limited number of A-BFT slots can be extended by the SA-BFT and the collision probability of A-BFT phase can be alleviated greatly, the number of A-BFT slots that can be extended is limited by the number of reserved bits (i.e., E-A-BFT Length) in the Beacon Interval Control field. The collision probability of beam training is still unsatisfactorily high in ultra-dense user scenarios. Once collision occurs, the A-BFT slot may be unavailable for BF training, which will result in great waste of BF training opportunities. In order to further alleviate the problem of high collision in ultra-dense user scenarios, we propose a solution named SBA-BFT. Firstly, we set a value $P$, $P\in(0,1]$ to constrain the number of EDMG STAs participating in an A-BFT phase. Before entering an A-BFT phase, each EDMG STA selects a random $p$, $p\in[0,P_j]$ where $j \in [0,n],{P_j} = 1 - j(1 - P)/{n}$ , $j$ is the number of times that EDMG STAs is prohibited from entering the A-BFT phase by $P$, and $n$ is the maximum prohibited times. The variation of $P_j$ is shown in Fig. 11. If $p \le P_j$, the STA is allowed to enter the A-BFT phase; otherwise, the STA is prohibited from entering the A-BFT phase. Thus, the probability of entering an A-BFT phase increases as the number of prohibited times increases. When an EDMG STA reaches the maximum prohibited times $n$, it will enter the A-BFT phase with the probability of 100\%. Then a secondary backoff mechanism is introduced into the second A-BFT slot region for EDMG STAs.

\begin{figure}[!htbp]
  \begin{center}
    \scalebox{0.8}[0.8]{\includegraphics{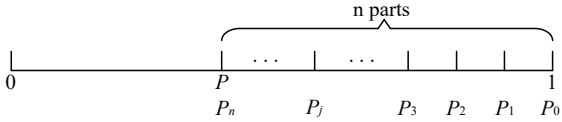}}
    \caption{Relationship between prohibited times $j$ and $P_j$.}
  \end{center}
\end{figure}

\subsection{SBA-BFT Mechanism}
As shown in Fig. 12, each EDMG STA allowed to enter an A-BFT phase shall randomly select an A-BFT slot and a random backoff timer within the A-BFT slot, namely secondary backoff. At the beginning of the selected A-BFT slot, instead of transmitting SSW frames to perform R-TXSS immediately, the EDMG STA starts the countdown to the secondary backoff phase. Only when the secondary backoff timer reaches zero, can the EDMG STA begin to transmit SSW frames. By employing the SBA-BFT mechanism, even if multiple EDMG STAs happen to select the same A-BFT slot, their secondary backoff timers may not necessarily the same. As a result, the EDMG STA with a shorter secondary backoff time will transmit SSW frames first, and the EDMG STA with a longer secondary backoff time will detect the channel is busy by doing Clear Channel Assessment (CCA) detection. Thus, it will not transmit SSW frames. In that case, the possible collision could be avoided.

\begin{figure}[!htbp]
  \begin{center}
    \scalebox{0.71}[0.71]{\includegraphics{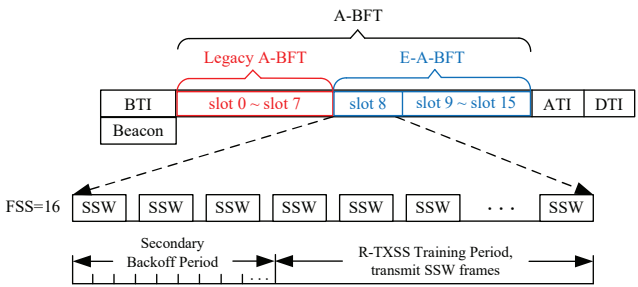}}
    \caption{Configuration of one A-BFT slot in the SBA-BFT.}
  \end{center}
\end{figure}

If one EDMG STA with an unsuccessful attempt in current A-BFT conflicts with other EDMG STAs in the next A-BFT, its BF training time will be delayed. In order to improve the timeliness of BF training for that EDMG STA, the secondary backoff window of the collided EDMG STAs should be shorter than that of newly joined EDMG STAs. Then the secondary backoff window decreases with the increase of the number of collided times of the EDMG STA.

\begin{figure*}[!htbp]
  \begin{center}
    \scalebox{0.64}[0.64]{\includegraphics{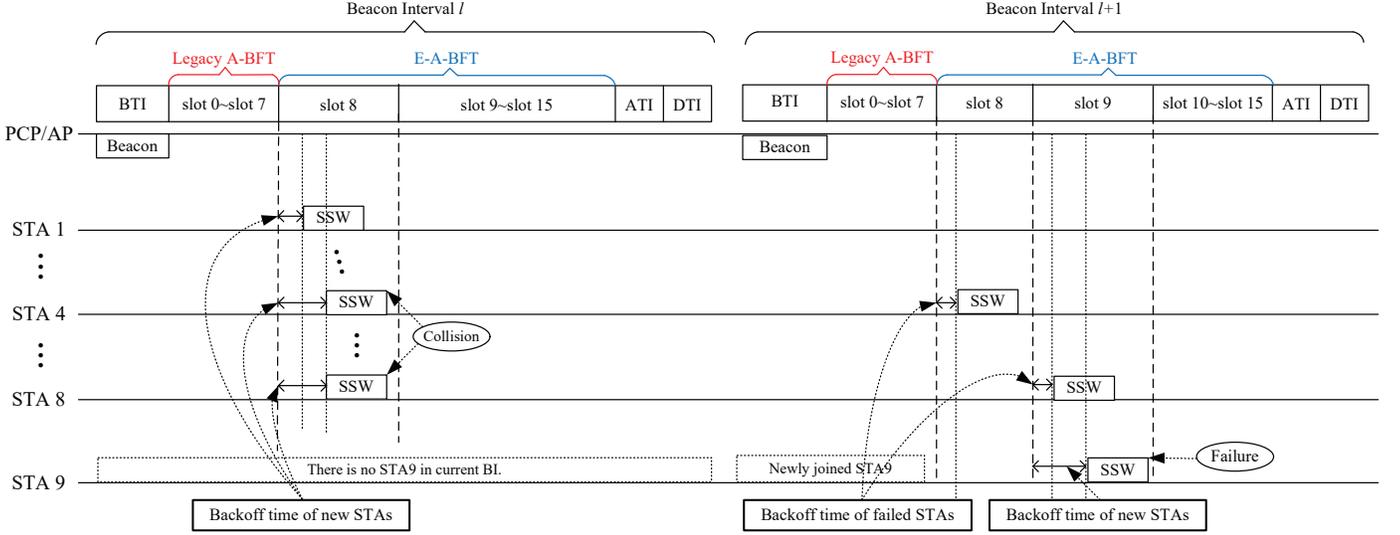}}
    \caption{An example of the SBA-BFT process.}
  \end{center}
\end{figure*}

Fig. 13 shows the diagram of an SBA-BFT process. In the $l$-$th$ BI, suppose EDMG STA 1, EDMG STA 4 and EDMG STA 8 select the same A-BFT slot (i.e., A-BFT slot 8) in the second A-BFT slot region. If they follow the IEEE 802.11ad A-BFT, these three EDMG STAs will transmit SSW frames simultaneously in A-BFT slot 8, which will definitely results in collision. However, in the proposed SBA-BFT, each of them will select a random secondary backoff timer. Suppose that the secondary backoff timer of EDMG STA 1 is shorter than that of EDMG STA 4 and EDMG STA 8, EDMG STA 1 will transmit SSW frames when its secondary backoff timer reaches zero. While EDMG STA 4 and EDMG STA 8 detect and get to know that the channel is busy by performing CCA, neither of them transmits SSW frames, and thus EDMG STA 1 successfully occupies A-BFT slot 8 for BF training. Assume that there is a newly joined EDMG STA 9 in the ($l$+1)-$th$ BI, it will compete with EDMG STA 4 and EDMG STA 8. Since EDMG STA 4 and EDMG STA 8 have failed in the $l$-$th$ BI, their secondary backoff window will be shorter than that of EDMG STA 9. If EDMG STA 8 and EDMG STA 9 select the same A-BFT slot (i.e., A-BFT slot 9), the secondary backoff timer of EDMG STA 8 will be more likely shorter than that of EDMG STA 9. Therefore, EDMG STA 8 will be more likely to occupy A-BFT slot 9 successfully. It is obvious that our SBA-BFT mechanism can fairly promote the priority of the failed EDMG STAs.

According to the complexity analysis in our previous work \cite{ref28}, assuming that the number of beams at the transmitter and receiver are $N_{TX}$ and $N_{RX}$, respectively. The complexity of the legacy 802.11ad is $O$($N_{TX}$*$N_{RX}$*`A-BFT Length'). Since the SA-BFT and SBA-BFT just extend additional `E-A-BFT Length' A-BFT slots, the complexity of SA-BFT and SBA-BFT are $O$($N_{TX}$*$N_{RX}$*`A-BFT Length + E-A-BFT Length').

\subsection{Secondary Backoff Window Design of the SBA-BFT}
Based on the analysis above, in order to promote the priority of the failed EDMG STAs, the secondary backoff window should decrease with the increase of failed times of an EDMG STA. We define the failed times of an EDMG STA as $i$, named as a backoff stage. According to the IEEE 802.11ad standard, we set the maximum number of failed times to be $m$ ($m < dot11RSSRetryLimit $). If it reaches the maximum number of failed times, the backoff window will no longer change. Therefore, if an EDMG STA fails $i$ times, its backoff window is ${W_i} = {2^{(m - i)}} \cdot W, i \in [0,m]$, where $W$ is the minimum backoff length (i.e., $aSlotTime$). We set the secondary backoff timer of a newly joined STA to $time_1$ which is randomly selected from $[0,W_0]$. The secondary backoff timer of an STA who fails $i$ times is set to $time_2$, which is randomly selected from $[0,{W_i}],i \in [0,m]$.

\subsection{Analytical Model for the SBA-BFT}
In this subsection, a three dimensional Markov chain is presented to model the SBA-BFT, which is based on the Markov model of Distributed Coordination Function (DCF) \cite{ref32}. However, there are several differences in our proposed Markov chain model. Firstly, the Markov model of DCF in \cite{ref32} does not contain the \emph{P} phase defined in Section IV to limit the number of STAs participating in A-BFT. Secondly, due to the existence of the \emph{P} phase, the transition between two backoff stages shall be determined by the probability of the corresponding \emph{P} phase (i.e., \emph{$P_j$}). Thirdly, the backoff window in our model is decreased with the number of failure times, while the backoff window in \cite{ref32} is increased with the number of failure times. Since all EDMG STAs randomly select an A-BFT slot in an A-BFT phase, the collision probabilities of every A-BFT slot can be approximately equal when there are a large number of contending EDMG STAs. For simplicity, this model focuses on one A-BFT slot, and some of the assumptions are listed as follows:

$a(t)$ is a stochastic process, and represents the number of times an EDMG STA is prohibited by $P$;
$b(t)$ is a stochastic process, and represents the secondary backoff timer for a given EDMG STA;
$s(t)$ is a stochastic process, representing the backoff stage of an EDMG STA at time $t$, where $t$ is a discrete integer;
$P_e$ is the success probability when an EDMG STA performs secondary backoff. $P_e$ can be considered as a fixed value and has no relationship to the backoff stage. The process ${\{a(t), b(t), s(t)}\}$ can be modeled as a three dimensional discrete-time Markov chain depicted in Fig. 14 \cite{ref32}, where ${b_{j,i,k}} = \mathop {\lim }\limits_{t \to \infty } P\{ a(t) = j,s(t) = i,b(t) = k\} ,j \in [0,n],i \in [0,m],k \in [0,{W_i} - 1]$.

\emph{Proposition 1}: The one step transition probabilities are given as
\begin{equation}
\left\{ \begin{array}{l}
P\{ j,i,k|j,i,k + 1\}  = 1, j = i, i \in [0,m], k \in [0,{W_i} - 2]\\
P\{ j,i,k|j,i,k + 1\}  = 1, j \in [m + 1,n], i = m, \\ \vspace{0.5em} k \in [0,{W_i} - 2]\\
P\{ j,i, - 1|j - 1,i - 1,0\}  = 1 - {P_e}, j = i, i \in [1,m]\\
P\{ j,i, - 1|j - 1,i,0\}  = 1 - {P_e}, j \in [m + 1,n], i = m\\
P\{ j,i, - 1|j,i,0\}  = 1 - {P_e}, j = n, i = m\\
P\{ 0,0, - 1|j,i,0\}  = {P_e}, j = i, i \in [0,m]\\
P\{ 0,0, - 1|j,i,0\}  = {P_e}, j \in [m + 1,n], i = m\\
P\{ j,i, - 1|j - 1,i - 1, - 1\}  = 1 - P/{P_{j - 1}}, j = i, \\ \vspace{0.5em} i \in [1,m]\\
P\{ j,i, - 1|j - 1,i, - 1\}  = 1 - P/{P_{j - 1}}, j \in [m + 1,n], \\ \vspace{0.5em} i = m\\
P\{ j,i,k|j,i, - 1\}  = \left( {P/{P_i}} \right)/{W_i}, j = i, i \in [0,m], \\ \vspace{0.5em} k \in [0,{W_i} - 1]\\
P\{ j,i,k|j,i, - 1\}  = \left( {P/{P_i}} \right)/{W_i}, j \in [m + 1,n], i = m, \\ \vspace{0.5em} k \in [0,{W_i} - 1].
\end{array} \right.
\end{equation}

\vspace{0.5em}

\emph{Proof}: The 1st and 2nd items in (1) stand for the fact that for a given A-BFT slot, the backoff timer is decremented. The 3rd to 5th items demonstrate the fact that if a collision occurs when the backoff timer reaches zero at backoff stage ($i$-1). In that case, the EDMG STA should go to the $i$-$th$ $P$ phase (i.e., failed $i$ times) and then randomly selects a $p$ to determine whether it can enter backoff stage $i$ or not. The 6th and 7th items stand for the fact that if an EDMG STA successfully performs R-TXSS training when the backoff time counter reaches zero at backoff stage $i$, this EDMG STA should go to the initial state (0, 0, -1). The 8th and 9th items stand for the fact that if an EDMG STA is prohibited again at $j$-$th$ $P$ phase, the EDMG STA should go to the ($j$+1)-$th$ $P$ phase. At last, the 10th and 11st items mean that if an EDMG STA succeeds at $j$-$th$ $P$ phase, the EDMG STA should go to the corresponding backoff stage $j$.

\begin{figure*}[!htbp]
  \begin{center}
    \scalebox{0.87}[0.87]{\includegraphics{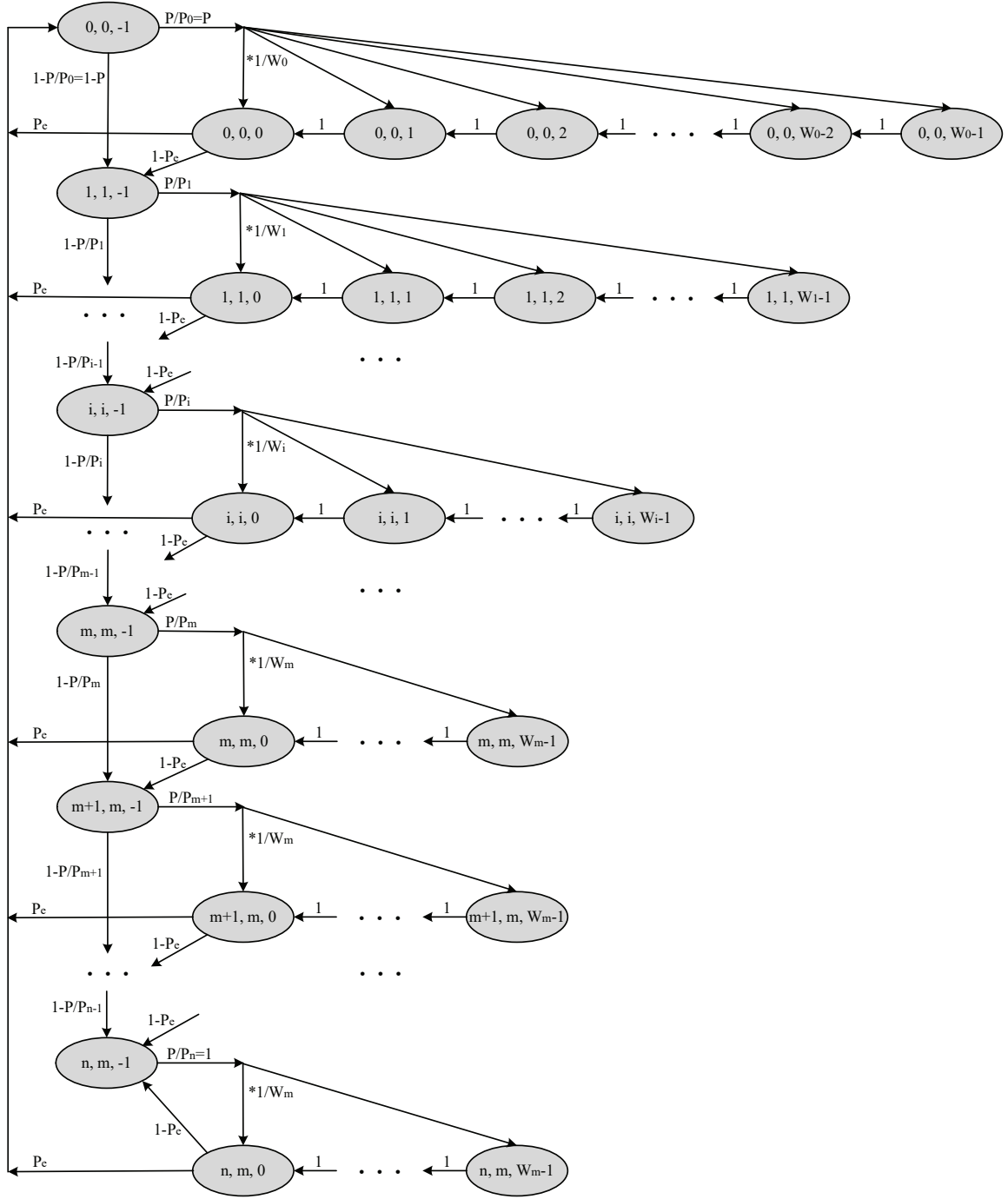}}
    \caption{Three-dimensional Markov model for the SBA-BFT.}
  \end{center}
\end{figure*}

To simplify the analysis, we assume that the maximum number of failed times at $P$ phase is equal to the maximum number of failed times at the secondary backoff phase (i.e., $m = n$). From the model, we observe that $j$ (the times prohibited by $P$) keeps in synchronization with $i$ (backoff stage), thus we can obtain that
\vspace{0.5em}
\begin{equation}
\begin{array}{l}
{b_{j,i,0}} = {b_{j - 1,i - 1,0}} \cdot \left( {1 - {P_e}} \right)\frac{P}{{{P_j}}} \\
 \to {b_{j,i,0}} = {\left( {1 - {P_e}} \right)^j} \cdot \frac{{{P^j}}}{{\prod\limits_{t = 1}^j {{P_t}} }} \cdot {b_{0,0,0}},j = i,i \in [1,m],
\end{array}
\end{equation}
and
\begin{equation}
{b_{j,i, - 1}} = \left\{ \begin{array}{l}
\vspace{0.5em}
\sum\limits_{t = 0}^m {{b_{t,t,0}}}  \cdot {P_e},j = i = 0 \\
{b_{j - 1,i - 1, - 1}} \cdot \left( {1 - \frac{P}{{{P_{j - 1}}}}} \right) + {b_{j - 1,i - 1,0}} \cdot \left( {1 - {P_e}} \right), \\ \vspace{0.5em} j = i,i \in (0,m) \\
{b_{j - 1,i - 1, - 1}} \cdot \left( {1 - \frac{P}{{{P_{j - 1}}}}} \right) + {b_{j - 1,i - 1,0}} \cdot \left( {1 - {P_e}} \right) + \\ \vspace{0.5em} {b_{j,i,0}} \cdot \left( {1 - {P_e}} \right),j = i = m.
\end{array} \right.
\end{equation}
\vspace{0.5em}

Owing to the chain regularities, for each $k \in [1,{W_i} - 1]$, we have
\vspace{0.5em}
\begin{equation}
{b_{j,i,k}} = \frac{{{W_i} - k}}{{{W_i}}} \cdot {b_{j,i, - 1}} \cdot \frac{P}{{{P_j}}},j = i,i \in [0,m].
\end{equation}
\vspace{0.5em}
Substituting (2) (3) into (4), ${b_{j,i,k}}$ can be rewritten as (5).
\begin{table*}[t]
\begin{equation}
{b_{j,i,k}}{\rm{ = }}\frac{{{W_i} - k}}{{{W_i}}} \cdot \frac{P}{{{P_j}}} \cdot {b_{0,0,0}\cdot}\left\{
\begin{array}{l}

\vspace{0.5em}

{P_e} + \sum\limits_{t = 1}^m {{{\left( {1 - {P_e}} \right)}^t} \cdot \frac{{{P^t}}}{{\prod\limits_{d = 1}^t {{P_d}} }} \cdot {P_e},j = i = 0} \\

\vspace{0.5em}

\left( {{P_e} + \sum\limits_{t = 1}^m {{{\left( {1 - {P_e}} \right)}^t} \cdot \frac{{{P^t}}}{{\prod\limits_{d = 1}^t {{P_d}} }} \cdot {P_e}} } \right) \cdot \left( {1 - \frac{P}{{{P_0}}}} \right) + \left( {1 - {P_e}} \right),j = i = 1\\

\left( {{P_e} + \sum\limits_{t = 1}^m {{{\left( {1 - {P_e}} \right)}^t} \cdot \frac{{{P^t}}}{{\prod\limits_{d = 1}^t {{P_d}} }} \cdot {P_e}} } \right) \cdot \left( {1 - \frac{P}{{{P_0}}}} \right) \cdot \left( {1 - \frac{P}{{{P_1}}}} \right) \\
\vspace{0.5em}
+ \left( {1 - {P_e}} \right) \cdot \left( {1 - \frac{P}{{{P_1}}}} \right)
 + \left( {1 - {P_e}} \right) \cdot \frac{P}{{{P_1}}} \cdot \left( {1 - {P_e}} \right), j = i = 2\\

\left( {{P_e} + \sum\limits_{t = 1}^m {{{\left( {1 - {P_e}} \right)}^t} \cdot \frac{{{P^t}}}{{\prod\limits_{d = 1}^t {{P_d}} }} \cdot {P_e}} } \right) \cdot \prod\limits_{d = 0}^{i - 1} {\left( {1 - \frac{P}{{{P_d}}}} \right)}  + \left( {1 - {P_e}} \right) \cdot \prod\limits_{t = 1}^{i - 1} {\left( {1 - \frac{P}{{{P_t}}}} \right)} \\
\vspace{0.5em}
 + \sum\limits_{t = 1}^{i - 2} {\left( {{{\left( {1 - {P_e}} \right)}^t} \cdot \frac{{{P^t}}}{{\prod\limits_{d = 1}^t {{P_d}} }} \cdot \left( {1 - {P_e}} \right) \cdot \prod\limits_{d = t + 1}^{i - 1} {\left( {1 - \frac{P}{{{P_d}}}} \right)} } \right)} \\
  + {\left( {1 - {P_e}} \right)^{i - 1}} \cdot \frac{{{P^{i - 1}}}}{{\prod\limits_{d = 1}^{i - 1} {{P_d}} }} \cdot \left( {1 - {P_e}} \right),j = i,i \in (2,m)\\

\left( {{P_e} + \sum\limits_{t = 1}^m {{{\left( {1 - {P_e}} \right)}^t} \cdot \frac{{{P^t}}}{{\prod\limits_{d = 1}^t {{P_d}} }} \cdot {P_e}} } \right) \cdot \prod\limits_{d = 0}^{i - 1} {\left( {1 - \frac{P}{{{P_d}}}} \right)}  + \left( {1 - {P_e}} \right) \cdot \prod\limits_{t = 1}^{i - 1} {\left( {1 - \frac{P}{{{P_t}}}} \right)} \\
 + \sum\limits_{t = 1}^{i - 2} {\left( {{{\left( {1 - {P_e}} \right)}^t} \cdot \frac{{{P^t}}}{{\prod\limits_{d = 1}^t {{P_d}} }} \cdot \left( {1 - {P_e}} \right) \cdot \prod\limits_{d = t + 1}^{i - 1} {\left( {1 - \frac{P}{{{P_d}}}} \right)} } \right)} \\
 + {\left( {1 - {P_e}} \right)^{i - 1}} \cdot \frac{{{P^{i - 1}}}}{{\prod\limits_{d = 1}^{i - 1} {{P_d}} }} \cdot \left( {1 - {P_e}} \right) + {\left( {1 - {P_e}} \right)^i} \cdot \frac{{{P^i}}}{{\prod\limits_{d = 1}^i {{P_d}} }} \cdot \left( {1 - {P_e}} \right),j = i = m.
\end{array} \right.
\end{equation}
\hrule
\end{table*}

Since $\sum\limits_{j = 0}^n {\sum\limits_{i = 0}^m {\sum\limits_{k =  - 1}^{{W_i} - 1} {{b_{j,i,k}}} } }  = 1$, where $n = m$, $j = i$ and ${W_i} = {2^{(m - i)}} \cdot W,i \in [0,m]$, we can obtain ${b_{0,0,0}}$, and it is only related to $P$, $P_e$, $m$, $W$ (where $P$, $m$, $W$ are constant values predetermined).

We can now express $p_{tr}$, the probability that an EDMG STA transmits in a randomly selected A-BFT slot. Any transmission will succeed when the backoff timer reaches zero, regardless of the backoff stage. That is
\vspace{0.5em}
\begin{equation}
{p_{tr}} = \sum\limits_{j = 0}^n {\sum\limits_{i = 0}^m {{b_{j,i,0}}} {\rm{ = }}{b_{0,0,0}} + \sum\limits_{i = 1}^m {{{\left( {1 - P{}_e} \right)}^i} \cdot \frac{{{P^i}}}{{\prod\limits_{d = 1}^i {{P_d}} }} \cdot {b_{0,0,0}}} ,}
\end{equation}
\vspace{0.5em}
where $n = m$, $j = i$. $p_{tr}$ can be expressed by ${b_{0,0,0}}$. Thus $p_{tr}$ is also related to $P$, $P_e$, $m$, $W$.

\emph{Proposition 2:} Assume that there are $s$ EDMG STAs involved in BF training in one A-BFT slot on average, the success probability when an EDMG STA performs a secondary backoff can be expressed as
\vspace{0.5em}
\begin{equation}
{P_e} = \sum\limits_{j = 1}^{{2^m} - 1} {\frac{{\left( {\begin{array}{*{20}{c}}
{\left\lceil {s \cdot {p_{tr}}} \right\rceil }\\
1
\end{array}} \right){{\left( {\begin{array}{*{20}{c}}
{{2^m} - 1 - j}\\
1
\end{array}} \right)}^{\left\lceil {s \cdot {p_{tr}}} \right\rceil  - 1}}}}{{{{\left( {{2^m}} \right)}^{\left\lceil {s \cdot {p_{tr}}} \right\rceil }}}}} .
\end{equation}
\vspace{0.5em}

\emph{Proof:} The proof is provided in Appendix A.

By means of (6) and (7), we can finally get the approximate solution $P_e$ for a given $s$.

\subsection{Optimization of the Maximum Number of Failed Times}
In the IEEE 802.11ad standard, since the minimum time of backoff is $W$ ($W$ = 5 \textmu s, $W$ equals to $aSlotTime$), the secondary backoff time of the SBA-BFT can be expressed as \cite{ref23}
\begin{equation}
\emph{Backoff Time} = \emph{Random()} \cdot aSlotTime.
\end{equation}

The maximum time of secondary backoff is
\begin{equation}
{T_{\max }} = {W_0} = {2^m} \cdot W = {2^m} \cdot 5 \mu s.
\end{equation}

The number of SSW frames can be transmitted in an A-BFT slot is indicated by the 4 bits \emph{FSS} field of the Beacon Interval Control element as shown in Fig. 6. It takes about 15 \textmu s (\emph{TXTIME(SSW)} = 15 \textmu s) to transmit one SSW frame in the DMG Control PHY in the IEEE 802.11ad standard \cite{ref33}. The interval between two SSW frames is a Short Beamforming Inter Frame Spacing (SBIFS), where SBIFS = 1 \textmu s. Therefore, when the maximum number of failed times is $m$, the maximum number of wasted SSW frames transmission opportunities is
\begin{equation}
\begin{array}{l}
{N_{waste}} = \left\lceil {\frac{{{T_{\max }}}}{{\emph{TXTIME(SSW)} + \emph{SIBFS}}}} \right\rceil  =
\left\lceil {\frac{{{2^m} \cdot 5}}{{15 + 1}}} \right\rceil = \left\lceil {5 \cdot {2^{m - 4}}} \right\rceil .
\end{array}
\end{equation}

Since the maximum number of SSW frames that can be transmitted in an A-BFT slot equals to 16 (i.e., \emph{FSS} = 16), the maximum value of $m$ can be set to 5.

In the IEEE 802.11ad standard, if two or more STAs select the same A-BFT slot, the success probability of this A-BFT slot is zero. However, when the SBA-BFT is adopted, the success probability of this A-BFT slot can be represented as (7). According to (7), we can infer that the success probability $P_e$ rises with the increase of $m$, whereas, the number of wasted opportunities which can be used to transmit SSW frames increases according to (10). Thus, we need to find an optimal value of $m$ to make a tradeoff between success probability and wasted BF training opportunities.

Assume that the success probability of an EDMG STA which performs the SBA-BFT is $P_e$, the number of SSW frames it can transmit is ${N_{send}} = 16 - {N_{waste}}$. On the other hand, the failure probability is $1-P_e$ , which means no frame can be transmitted. Thus in the SBA-BFT, the number of SSW frames can be transmitted within one A-BFT slot can be specified as
\begin{equation}
{N_{slot}} = {N_{send}} \cdot {P_e} + 0 \cdot [1 - {P_e}] = \left( {16 - {N_{waste}}} \right) \cdot {P_e},
\end{equation}
where $ s>1, m \le 5 $.

Therefore, the optimal value of $m$ is
\begin{equation}
m = \arg \max {N_{slot}} = \arg \max \left( {16 - \left\lceil {5 \cdot {2^{m - 4}}} \right\rceil } \right) \cdot {P_e}.
\end{equation}

\subsection{Overload Indicator Definition}
 If there are a very small number of EDMG STAs, adopting the SBA-BFT will waste the opportunities of transmitting SSW frames in one A-BFT slot. We intend to use the reserved bit B44 (as shown in Fig. 6) of Beacon Interval Control element as the A-BFT Overload Indicator (OI). PCP/AP can estimate the number of EDMG STAs involved in A-BFT based on historical BF training information \cite{ref23}, \cite{ref34}. For example, if there are plenty of STAs participating in A-BFT training contention in the previous A-BFT phase, the interference and noise detected by PCP/AP will be at a high level (can be realized through CCA function at PCP/AP side), and if the number of failure times \emph{i} is greater than a predefined threshold, this A-BFT phase can be considered as overloaded. Once there are a few STAs participating in A-BFT training in the previous A-BFT phase, the interference and noise detected by PCP/AP will be at a lower level. We define two parameters $S$ and $N_{th}$ to respectively represent the number of STAs involved in A-BFT and the threshold which will indicate whether A-BFT is overloaded or not. If there are a few EDMG STAs involved in A-BFT, which means $S < N_{th}$, it will be inefficient to adopt the SBA-BFT. In that case, PCP/AP can set OI to 0, which means only the legacy A-BFT phase will be performed. Otherwise, OI shall be set to 1, thus the SBA-BFT shall be executed to alleviate the serious collision. The SBA-BFT algorithm is shown in Algorithm 2 (since $j$ keeps in synchronization with $i$ and they both stand for the failed times, we use $j$ to replace $i$ in Algorithm 2).

\begin{algorithm}[!ht]
  \caption{SBA-BFT Algorithm.}
  \begin{algorithmic}[1]
   \STATE PCP/AP estimates the overload level of A-BFT based on historical BF training information.
   \STATE If $S < N_{th}$, set OI = 0. The legacy IEEE 802.11ad A-BFT shall be applied during A-BFT phase;
   \STATE If $S \ge N_{th}$, set OI = 1. A-BFT phase adopts the SBA-BFT. PCP/AP assigns the values of $P$, `A-BFT Length' and `E-A-BFT Length'. The Start Time field is set to `A-BFT Length + E-A-BFT Length'.
   \STATE Set $j$ to 0, and ${W_j} = {2^{(m - j)}} \cdot W, j \in [0,m]$, where $m$ is the maximum failed times.
    \STATE PCP/AP transmits DMG Beacon frames to perform I-TXSS training.
    \STATE DMG STAs determine their A-BFT slot region according to `A-BFT Length' field of DMG Beacon frame and perform R-TXSS training in the region by adopting IEEE 802.11ad A-BFT.
    \STATE Each EDMG STA randomly selects a $p$ from $[0,P_j]$, where $P_j = 1 - j\frac{{1 - P}}{n},j \in [0,n],$
    if $p \le P_j$, it is allowed to enter A-BFT phase;
    if $p > P_j$, it is prohibited to enter the A-BFT phase, $j = j + 1$ and redo step 7 in the next BI.
    \vspace{-1em}
    \STATE The allowed EDMG STAs determine their A-BFT slot region according to `A-BFT Length' field and `E-A-BFT Length' field. The start point is `A-BFT Length' and region length is `E-A-BFT Length'.
    \STATE For a newly joined EDMG STA, it randomly selects a secondary backoff counter $time_1$ from [0,$W_0$] and for an EDMG STA who failed $j$ times, it randomly selects a secondary backoff counter $time_2$ from [0,$W_j$] .
    \STATE When the secondary backoff time counter reaches zero, if the channel is sensed to be idle via CCA, the EDMG STA will transmit SSW frames to perform R-TXSS training and it will become a newly joined STA in the next BI. If the channel is sensed to be busy, the EDMG STA will not transmit SSW frames, $j = j + 1$ and the EDMG STA will redo step 7 in the next BI.
  \end{algorithmic}
\end{algorithm}

\section{Performance Evaluation}
In this section, extensive simulations have been carried out to evaluate the performance of the SA-BFT and SBA-BFT. First of all, we demonstrate the success probability of single slot for the three schemes.. The related simulation parameters are chosen based on the IEEE 802.11ad and 802.11ay standards \cite{ref23}, \cite{ref33}, and are listed in Table II.

\begin{table}[!htbp]
\normalsize
\caption{Simulation Parameters I}
\centering
\begin{tabular}{|c|c|}
\hline
\textbf{Parameters} & \textbf{Values} \\
\hline
\emph{m} & 1, 3 \\
\hline
\emph{s} & [1,10] \\
\hline
\emph{FSS} & 16 \\
\hline
E-A-BFT Length & 8 \\
\hline
\emph{aSlotTime} & 5 \textmu s \\
\hline
\emph{TXTIME(SSW)} & 15 \textmu s \\
\hline
\emph{SBIFS} & 1 \textmu s \\
\hline
\end{tabular}
\end{table}

\begin{figure}[!htbp]
  \begin{center}
    \scalebox{0.49}[0.49]{\includegraphics{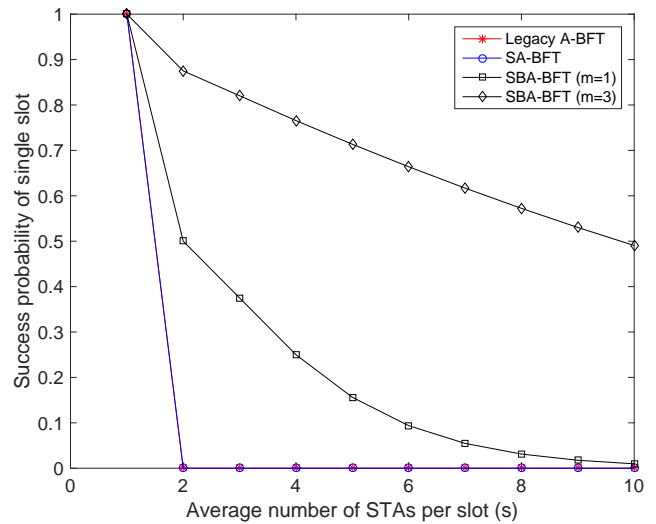}}
    \caption{Success probability of single slot for the three schemes.}
  \end{center}
\end{figure}

For the SBA-BFT working in an A-BFT phase, even multiple EDMG STAs happen to select the same A-BFT slot, collision shall occur only when they select the same secondary backoff time. As illustrated in Fig. 15, the success probability of $m$ = 3 is higher than that of $m$ = 1 since there are more space for secondary backoff. On the other hand, if the legacy IEEE 802.11ad A-BFT and the SA-BFT are working properly, collisions occur under the condition that there are two or more EDMG STAs selecting the same A-BFT slot. It is observed that the SBA-BFT can significantly improve the utilization of A-BFT slots in dense user scenarios. In other words, the number of successfully trained EDMG STAs can be increased.

\begin{table}[!htbp]
\normalsize
\caption{Simulation Parameters II}
\centering
\begin{tabular}{|c|c|}
\hline
\textbf{Parameters} & \textbf{Values} \\
\hline
\emph{m} & 3 \\
\hline
\emph{S} & [1,30] \\
\hline
\emph{FSS} & 16 \\
\hline
A-BFT Length & 8 \\
\hline
E-A-BFT Length & 8 \\
\hline
\emph{aSlotTime} & 5 \textmu s \\
\hline
\emph{TXTIME(SSW)} & 15 \textmu s \\
\hline
\emph{SBIFS} & 1 \textmu s \\
\hline
\end{tabular}
\end{table}

Furthermore, the SA-BFT is able to provide more A-BFT slots, thus increases the number of successful EDMG STAs. The simulation parameters are listed in Table III. Note that parameter $S$ stands for the total number of STAs, which consist of both DMG STAs and EDMG STAs. Fig. 16 shows the number of successful STAs comparison among the IEEE 802.11ad A-BFT, the SA-BFT and the SBA-BFT. We can see from Fig. 16 that with the increase number of contending STAs, the legacy IEEE 802.11ad A-BFT achieves the maximum number of successful STAs when there are about 8 contending STAs. As the number of contending STAs increases, the number of successful STAs decreases in the IEEE 802.11ad A-BFT. Since the SA-BFT can provide more A-BFT slots than the legacy A-BFT scheme, the number of successful STAs could be 6. When there are more than 5 contending STAs, the SA-BFT can alleviate collisions greatly. As expected, the SBA-BFT promotes the number of successful STAs significantly compared to the legacy A-BFT and the SA-BFT, especially in dense user scenarios (i.e., more than 20 contending STAs). Furthermore, it is possible that an A-BFT slot selected by two or more STAs can still be successfully used for BF training by employing the SBA-BFT.

\begin{figure}[!htbp]
  \begin{center}
    \scalebox{0.49}[0.49]{\includegraphics{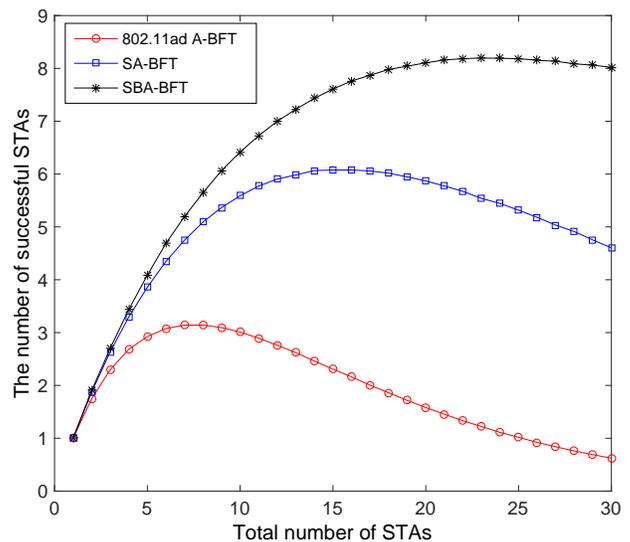}}
    \caption{The number of successful STAs comparison among the three mechanisms.}
  \end{center}
\end{figure}

It is worth mentioning that, the SBA-BFT performs secondary backoff at the expense of sacrificing a small number of transmit opportunities of SSW frames. Furthermore, the SBA-BFT is not so efficient in the case of sparse user scenarios. The following simulation investigates the condition to trigger the SBA-BFT (i.e., the value of OI). The simulation parameters are the same as in Table III.

\begin{figure}[!htbp]
  \begin{center}
    \scalebox{0.5}[0.5]{\includegraphics{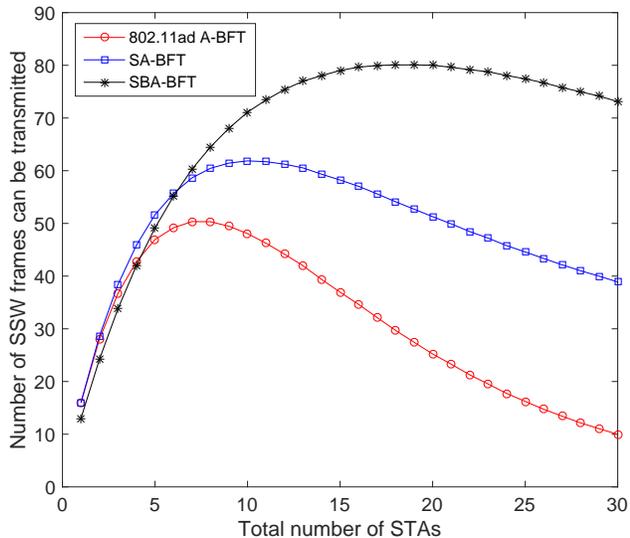}}
    \caption{The number of SSW frames can be transmitted.}
  \end{center}
\end{figure}

Fig. 17 demonstrates that the performance of the SBA-BFT is lower than that of the legacy IEEE 802.11ad A-BFT in sparse user scenarios. The reason lies that some transmission opportunities may be wasted with low collision probability. While in dense user scenarios (i.e., more than 5 STAs), the SBA-BFT has a significant performance improvement compared to the legacy IEEE 802.11ad A-BFT. Therefore, we can infer that the optimal value of $N_{th}$ could be 6. In this way, we can make sure that the performance will not decrease in sparse user scenarios because the SBA-BFT will not be triggered. Meanwhile, the performance will be significantly improved in dense user scenarios. Because the SBA-BFT promotes the success probability of one A-BFT slot when multiple STAs compete simultaneously. With the help of our proposed SA-BFT and SBA-BFT, the collision of BF training can be alleviated, and the success probability of A-BFT slots can be improved. Moreover, the BF training process will be faster in dense user scenarios.

The number of antennas and possible beamwidth will have impact on the choice of beam training methods and there are many on this issue. For example, \cite{ref35} analyzed the complexity of different beam training methods with different number of antenna elements and beamwidth. Then, it proposed a novel pre-search algorithm to reduce the beam training overhead. Impacts of beamwidth on throughput and beamforming training overhead are discussed in \cite{ref36}. There are also some detailed discussions on beam training time and beam training methods with different kinds of beamwidths (e.g., from 1$^{\circ}$ to 10$^{\circ}$) in \cite{ref37}. Similar results can also be found in \cite{ref38}, etc. In the 802.11ad and 802.11ay standards, with more antennas and narrower beams involved, there will be more time consumed for beam training. Therefore, efficient beam training methods are needed to reduce the time for beam training. Since the SA-BFT and SBA-BFT can improve the successful probability of beam training in A-BFT phase which means they can accelerate the beam training processes. Thus, the SA-BFT and SBA-BFT should be adopted instead of legacy A-BFT in massive MIMO beam training. Furthermore, the SBA-BFT outperforms the SA-BFT in ultra-dense user scenarios. Thus, the SBA-BFT should be considered first in the beam training of ultra-dense user scenarios. In this paper, we have not discussed massive MIMO based on the fact that Wireless Fidelity (Wi-Fi) aims to provide high quality of services with low cost and low complexity. We believe massive MIMO is too complicated to be considered for the current 802.11ad and the 802.11ay.

\section{Conclusion}

For future co-existance of the IEEE 802.11ad and IEEE 802.11ay mmWave networks, we propose an enhanced random access and BF training mechanism to alleviate high collision probability and low BF training efficiency problems in A-BFT phase in dense user scenarios. By employing the SA-BFT, we can provide more A-BFT slots for EDMG STAs to compete, which can alleviate collisions greatly. Besides, the SA-BFT can divide the A-BFT slots into two non-overlapping regions. The legacy DMG STAs compete for the first region and the EDMG STAs compete for the second region, and then EDMG STAs can perform the SBA-BFT in the second region and maintain compatibility with the IEEE 802.11ad standard as well. By performing secondary backoff, the SBA-BFT can further improve the BF training efficiency in dense user scenarios. The proposed backoff window transformation method can promote the priorities of failed EDMG STAs, thus improving the timeliness of BF training and the system quality of experience (QoE).

Theoretical and simulation results have verified that the proposed SA-BFT and SBA-BFT can not only increase the number of successful STAs, but also increase the number of transmitted SSW frames. It is also obvious that the proposed mechanisms can improve the BF training efficiency. The proposed schemes are expected to effectively handle user dense scenarios for future mmWave wireless communication systems.

\appendices
\section{Proof of proposition 2}

Different from the way to solve the collision probability of the IEEE 802.11 DCF \cite{ref32}, in the SBA-BFT when we set the maximum failure times to $m$, one A-BFT slot will be divided into $2^m$ subslots (one subslot equals to $aSlotTime$, i.e., the minimum backoff time $W$, so the maximum backoff window is ${2^m} \cdot W$, the collision probability is uncorrelated with $W$), as long as the first selected subslot in an A-BFT slot is only selected by one EDMG STA. The subsequent EDMG STAs will be aware of the channel busyness through CCA detection. They will not transmit if the channel is busy. Thus, the collision is avoided and this A-BFT slot is a successful slot for BF training. Since there will be $\left\lceil {s \cdot {p_{tr}}} \right\rceil $ EDMG STAs contending for the same A-BFT slot, the success only occurs when the first selected subslot $j, j \in [0,{2^m} - 1)$ is only selected by one EDMG STA and no EDMG STA selects subslots from subslot 0 to subslot $j$-1 and the subsequent $\left\lceil {s \cdot {p_{tr}}} \right\rceil - 1$ EDMG STAs select subslots from subslot $j$ + 1 to subslot ${2^m} - 1$. Thus, there are ${2^m} - 1$ kinds of success conditions. First, we can select one EDMG STA from $\left\lceil {s \cdot {p_{tr}}} \right\rceil $ EDMG STAs to get subslot $j$, where there are
$\left(
{\begin{array}{*{20}{c}}
{\left\lceil {s \cdot {p_{tr}}} \right\rceil }\\ 1
\end{array}}
\right)$
kinds of choices. Then the rest $\left\lceil {s \cdot {p_{tr}}} \right\rceil - 1$ EDMG STAs select the subslots from subslot $[j + 1,{2^m} - 1]$, where there are ${\left( {\begin{array}{*{20}{c}}
{{2^m} - 1 - j}\\
1
\end{array}} \right)^{\left\lceil {s \cdot {p_{tr}}} \right\rceil  - 1}}$ kinds of choices. Since there are ${\left( {{2^m}} \right)^{\left\lceil {s \cdot {p_{tr}}} \right\rceil }}$ kinds of choices in total, the success probability when EDMG STAs perform secondary backoff is ${P_e} = \sum\limits_{j = 1}^{{2^m} - 1} {\frac{{\left( {\begin{array}{*{20}{c}}
{\left\lceil {s \cdot {p_{tr}}} \right\rceil }\\
1
\end{array}} \right){{\left( {\begin{array}{*{20}{c}}
{{2^m} - 1 - j}\\
1
\end{array}} \right)}^{\left\lceil {s \cdot {p_{tr}}} \right\rceil  - 1}}}}{{{{\left( {{2^m}} \right)}^{\left\lceil {s \cdot {p_{tr}}} \right\rceil }}}},\left\lceil {s \cdot {p_{tr}}} \right\rceil  > 1} $. If there is only one EDMG STA to perform the SBA-BFT (i.e., $\left\lceil {s \cdot {p_{tr}}} \right\rceil  = 1$, there is no EDMG STA competes with it), the success probability is $P_e$ = 1.


\ifCLASSOPTIONcaptionsoff
  \newpage
\fi


\end{document}